\documentclass{optica-article}

\journal{opticajournal} % for journals or Optica Open

\articletype{Research Article}

\usepackage{lineno}
\usepackage{makecell}

\usepackage{multirow}
\usepackage{graphicx}

\usepackage{xcolor}

\begin{document}

\title{High-Efficiency InGaP-on-Insulator Microresonator Nonlinear Conversion and Entanglement Generation}

\author{
Xuefeng Li,\authormark{1,*}
Lillian Thiel,\authormark{1}
Yiming Pang,\authormark{1}
Amalu Shimamura,\authormark{1}
Lucas Wang,\authormark{2}
Joshua E. Castro,\authormark{1}
Max Meunier,\authormark{1}
Nicholas Lewis,\authormark{1}
%\newline
John E. Bowers,\authormark{1}
Kevin L. Silverman,\authormark{3}
Richard P. Mirin,\authormark{1}
Galan Moody\authormark{1,*}}

\address{
\authormark{1}Department of Electrical and Computer Engineering, University of California, Santa Barbara, CA, USA\\
\authormark{2}Department of Physics, University of California, Santa Barbara, CA, USA10\\
\authormark{3}National Institute of Standards and Technology, Boulder, CO, USA}

\email{\authormark{*}xuefengli@ucsb.edu} %% email address is required; see note below about the corresponding author designation
\email{\authormark{*}moody@ucsb.edu} %% email address is required; see note below about the corresponding author designation

\begin{abstract*} 
InGaP-on-insulator (InGaP-OI), with its intrinsically high $\chi^{\left(2\right)}$ optical nonlinearity, has emerged as an efficient and bright integrated photonic platform for frequency conversion and on-chip entanglement generation, but high waveguide propagation loss in the visible wavelength range has limited its overall performance. Here, we identify the dominant loss mechanism through mode-profile analysis and effectively mitigate the loss using a surface treatment method. Statistical analysis of the resonator quality factor and propagation loss reveals the optimal ring radius that maintains a strong nonlinear interaction while suppressing significant bending related loss, resulting in loss as low as 0.49 dB/cm (4.31 dB/cm) at 1560 nm (780 nm). The method provides a 3.5--4$\times$ quality factor enhancement at 780 nm, enabling a second-harmonic generation efficiency of $3.01\times10^{5}$ \,\%/\textrm{W} and a degenerate photon-pair generation rate of $4.27\,\textrm{MHz}/\mu\textrm{W}$ and coincidence-to-accidental ratio as high as 10,000. The quasi-phase matching condition is experimentally verified, and nonlinear conversion is systematically characterized across the entire parameter space. This work establishes a scalable pathway for classical and quantum photonics in a low-loss, highly nonlinear, and wafer-scale integration platform. 

\end{abstract*}

%%%%%%%%%%%%%%%%%%%%%%%%%%  body 
\section{Introduction}
Integrated nonlinear photonics plays an important role in scalable classical and quantum optical technologies. On-chip second-order ($\chi^{(2)}$) nonlinear processes enable efficient frequency conversion~\cite{yang2007enhanced,kuo2011and,zhao2022ingap,akin2024ingap,mckenna2022ultra,bruch201817,ahler2026low}, entangled photon-pair generation~\cite{zhao2022ingap,akin2024ingap,guo2017parametric,chopin2023ultra}, and optical information processing, such as squeezed light~\cite{vahlbruch2016detection}, through difference-frequency generation (DFG), second-harmonic generation (SHG), and spontaneous parametric down-conversion (SPDC). Significant progress has also been achieved in $\chi^{(3)}$-based integrated photonic platforms for quantum light generation, four-wave-mixing-based nonlinear optics, and frequency-comb generation~\cite{zeng2024quantum,chen2024ultralow,zeng2026integrated}. A nonlinear photonic platform that simultaneously offers high nonlinear efficiency, low propagation loss, and compatibility with large-scale heterogeneous integration remains in high demand for practical classical and quantum photonic systems.

Several $\chi^{\left(2\right)}$ systems have produced promising results. Thin-film lithium niobate (TFLN) is one leading platform due to its low propagation loss and high electro-optic efficiency~\cite{zhu2021integrated}. However, challenges remain in heterogeneous integration and monolithic incorporation with active photonic components. Wide-bandgap III--V semiconductors provide a compelling alternative. In particular, indium gallium phosphide (InGaP) is attractive due to its large second-order nonlinear coefficient ($\chi^{(2)} \approx 220\,\mathrm{pm/V}$~\cite{ueno1997second,ahler2026low}) and wide bandgap ($\sim1.9\,\mathrm{eV}$), which enables visible wavelength operation and eliminates two-photon absorption at infrared wavelengths. Its high refractive index enables tight modal confinement and strong nonlinear interaction. Furthermore, wafer-scale heterogeneous bonding has been demonstrated on a 4-inch InGaP epitaxial wafer, providing a pathway toward scalable InGaP-on-insulator (InGaP-OI) photonic integration~\cite{thiel2024wafer}.

Cavity enhancement in microresonators dramatically increases the nonlinear interaction strength by increasing intracavity field intensity and enabling resonant buildup of the interacting modes. In $\chi^{(2)}$ systems, doubly resonant microresonators significantly reduce pump power requirements and enable efficient SHG and SPDC in ten-micron-scale compact devices~\cite{zhao2022ingap,bruch201817}. However, a major challenge in III--V platforms, including InGaP-OI, is the significant propagation loss at visible wavelengths~\cite{zhao2022ingap,akin2024ingap}. Surface scattering from epitaxial interfaces, together with absorption associated with native oxides and surface states, can severely limit cavity quality factors. These effects become more pronounced at visible wavelengths, where scattering loss increases rapidly and the photon energy approaches the material band edge~\cite{gao2022probing,khalatpour2025roughness}. Although efficient SHG and SPDC have been demonstrated in small-radius, suspended InGaP microresonators, mechanical stability, thermal stability, and large-scale photonic integration remain fundamentally challenging~\cite{zhao2022ingap,akin2024ingap}. Therefore, realizing low-loss, high-performance classical and quantum nonlinear photonics in heterogeneously integrated InGaP-OI platforms remains a significant challenge.

In this work, we demonstrate high-efficiency nonlinear conversion and entanglement generation in low-loss InGaP-OI microresonators. Through stepper-based wafer-scale fabrication, systematic identification and mitigation of dominant loss mechanisms, as illustrated in Fig.~\ref{fig:A1}(a), and dispersion-engineered resonator design, we achieve an SHG efficiency of $3.01\times10^{5}\,\%/\mathrm{W}$ and a degenerate photon-pair generation rate of $4.27\,\mathrm{MHz}/\mu\mathrm{W}$ in $20\,\mu\mathrm{m}$-radius microresonators. These results establish simultaneous high-performance classical and quantum nonlinear processes within a monolithically integrated InGaP-OI platform. Our work positions InGaP-OI microresonators as scalable building blocks for next-generation frequency converters and integrated quantum photonic circuits.

% \begin{figure}[tbp]
% \centering\includegraphics[width=\textwidth]{Opt_comparison.pdf}
% \caption{
% (a) Comparison of broadband second-harmonic generation performance across nonlinear thin-film platforms, showing 
% absolute conversion efficiency (in \%), 
% normalized efficiency (in W$^{-1}$),
% and nonlinear strength 
% (in cm$^{-2}$/W).
% Data are compiled from 
% Refs.~\cite{Fang_InGaP, GaP2, best_PPLN_adapted, PPLT, best_PPLN_no_adapt, high_abs_PPLN, OP_LN, Placke_GaAs, Ulsig_GaAs, Stanton_GaAs, AOP_SiN_dual_resonance}. 
% Abbreviations include orientation-patterned (OP) and all-optical poling (AOP). 
% $^{a)}$Pumped at 1950 nm. 
% $^{b)}$Not pumped to depletion. 
% $^{c)}$Resonantly enhanced. 
% (b) Photograph of a full 100 mm InGaP-OI wafer.
% (c) Optical micrograph of a spiral waveguide used for SPDC,
% with the crystal axes indicated.
% (d) Optical micrographs of spiral and directional-coupler structures used for device characterization.}
% \label{fig:comparison}
% \end{figure}

\section{Nonlinear Interactions in Microresonators}
To enhance the nonlinear interaction between the pump and generated resonances, we use microresonators to build up energy at both wavelengths. The single-photon mode coupling coefficient for a resonator made of disordered InGaP~\cite{delong1995band} is given by~\cite{luo2019optical}:

\begin{equation}
g = \sqrt{\frac{\hbar \omega_{\mathrm{IR}}^{2}\omega_{\mathrm{vis}}}{8\epsilon_0}}
\,
\frac{
\int d r \,
\chi_{xyz}^{(2)}
\sum_{i\neq j\neq k}
E_{\mathrm{IR},i}^{*}
E_{\mathrm{IR},j}^{*}
E_{\mathrm{vis},k}
}{
\int dr\,\epsilon_{r,\mathrm{IR}} |E_{\mathrm{IR}}|^{2}
\sqrt{\int dr\,\epsilon_{r,\mathrm{vis}} |E_{\mathrm{vis}}|^{2}}
} .
\end{equation}

\noindent Here, $\hbar$ is the Planck constant. $\omega_{\mathrm{IR}}$ and $\omega_{\mathrm{vis}}$ are the angular frequencies of the infrared (IR) and visible (VIS) resonances, respectively. $E_{\mathrm{IR}}$ and $E_{\mathrm{vis}}$ are the electric fields of the IR and VIS modes. $\epsilon_0$, $\epsilon_{r,\mathrm{IR}}$, and $\epsilon_{r,\mathrm{vis}}$ are the vacuum permittivity and the relative permittivities at IR and VIS frequencies.

In addition to a large modal overlap, efficient nonlinear interaction requires the resonator to simultaneously satisfy energy conservation and quasi-phase matching. Energy conservation between IR ($\sim1560\,\mathrm{nm}$ ) and VIS ($\sim780\,\mathrm{nm}$) requires

\begin{equation}
\omega_{\mathrm{vis}} = 2\omega_{\mathrm{IR}} .
\end{equation}

\begin{figure}[h]
\centering\includegraphics[width=\textwidth]{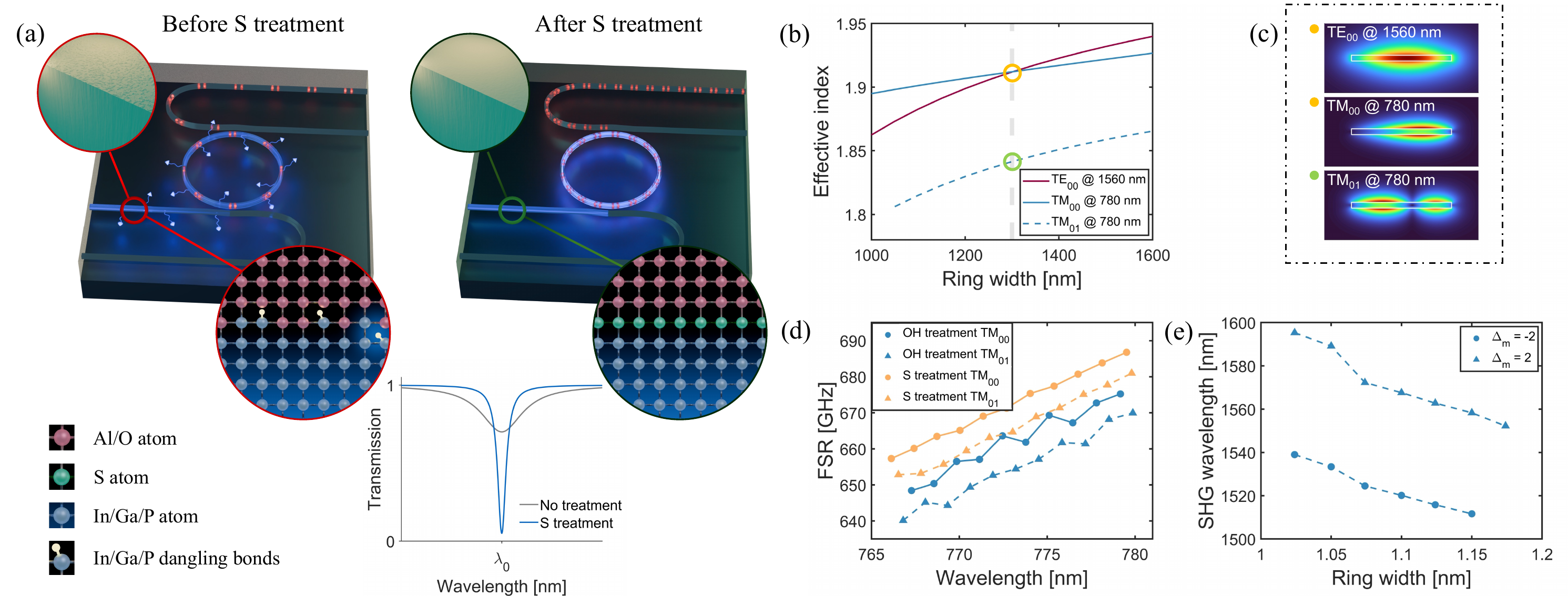}
\caption{
(a) Depiction of the effect of S treatment on surface roughness and surface states. For the same resonator configuration, the device with S treatment exhibits critical coupling, whereas the untreated device is significantly undercoupled and shows much higher propagation loss. The upper-left insets are schematic illustrations (not SEM images) of the InGaP surface before and after sulfur treatment. (b) FDE simulated effective index of the $1560\,\mathrm{nm}$ $\mathrm{TE}_{00}$ mode and the $780\,\mathrm{nm}$ $\mathrm{TM}_{00}$ and $\mathrm{TM}_{01}$ modes. (c) Simulated mode profiles of the $1560\,\mathrm{nm}$ $\mathrm{TE}_{00}$ mode and the $780\,\mathrm{nm}$ $\mathrm{TM}_{00}$ and $\mathrm{TM}_{01}$ modes. (d) Measured free spectral range (FSR) as a function of wavelength for TM modes with different surface treatments. (e) Measured SHG wavelength versus ring width for $\Delta m=\pm2$.}
\label{fig:A1}
\end{figure}

In zincblende-structured crystal resonators with $\overline{4}3m$ symmetry, quasi-phase matching can be expressed in terms of the azimuthal mode numbers of the interacting modes, and the strongest nonlinear coupling occurs when

\begin{equation}
\Delta m = 2m_{\mathrm{IR}} - m_{\mathrm{vis}} = \pm 2 ,
\end{equation}

\noindent where $m_{\mathrm{IR}}$ and $m_{\mathrm{vis}}$ correspond to the IR and VIS resonances, respectively~\cite{yang2007enhanced,kuo2011and}. The absolute values of the azimuthal mode numbers are difficult to determine experimentally because they are very large ($m_{\mathrm{IR}}\sim150$ and $m_{\mathrm{vis}}\sim300$) in $20\,\mu\mathrm{m}$-radius resonators. However, these large mode numbers indicate that quasi-phase matching roughly aligns with modal phase matching conditions~\cite{chang2019strong}. From finite difference eigenmode (FDE) simulations, a $105\,\mathrm{nm}$ thick In$_{0.5}$Ga$_{0.5}$P resonator with a $1300\,\mathrm{nm}$ ring width is expected to support modal phase matching between the $1560\,\mathrm{nm}$ TE$_{00}$ mode and the $780\,\mathrm{nm}$ TM$_{00}$ mode in a $20\,\mu\mathrm{m}$-radius ring resonator. The simulated effective index versus ring width for these modes is shown in Fig.~\ref{fig:A1}(b). For comparison, the $780\,\mathrm{nm}$ TM$_{01}$ mode is also included, which exhibits a much lower effective index within the studied ring-width range.

The corresponding mode profiles, including $\mathrm{TE}_{00}$ ($1560\,\mathrm{nm}$), $\mathrm{TM}_{00}$ ($780\,\mathrm{nm}$), and $\mathrm{TM}_{01}$ ($780\,\mathrm{nm}$), are shown in Fig.~\ref{fig:A1}(c). At $1560\,\mathrm{nm}$, the $\mathrm{TE}_{00}$ mode is largely confined within the waveguide core, resulting in relatively weak interaction with the top and bottom surfaces and even weaker interaction with the sidewalls. In contrast, the $780\,\mathrm{nm}$ TM modes exhibit strong overlap with the top and bottom surfaces, while the interaction with the waveguide core and sidewalls is noticeably weaker. For the TM modes, this represents a typical low-confinement waveguide configuration, usually realized using high-aspect-ratio geometries to minimize the interaction between the optical modes and the waveguide sidewalls and thereby improve the resonator quality factor~\cite{spencer2014integrated,jin2021hertz}. In such thin waveguides, scattering from the sidewalls becomes less significant, while the top and bottom surfaces play a critical role in determining the overall optical loss. As a result, optical loss at $780\,\mathrm{nm}$ is highly sensitive to surface roughness and is also strongly influenced by III--V surface states, which are known to degrade device performance in electronic and optoelectronic systems~\cite{wong2023recovering,hasegawa2010surface,jeng2007effective}. In III--V semiconductors, surface dangling bonds generate surface states that introduce electronic energy levels within the bulk semiconductor bandgap and can facilitate nonradiative carrier recombination while interacting with optical fields and carriers through absorption and defect-assisted recombination processes~\cite{brennan2011optimisation,laukkanen2021passivation,li2024multiple}. This phenomenon, commonly referred to as surface recombination, has been extensively studied in electronic and optoelectronic devices and is strongly associated with the surface-to-volume ratio of the active region~\cite{coldren2012diode}.

\begin{figure}[t]
\centering\includegraphics[width=0.7\columnwidth]{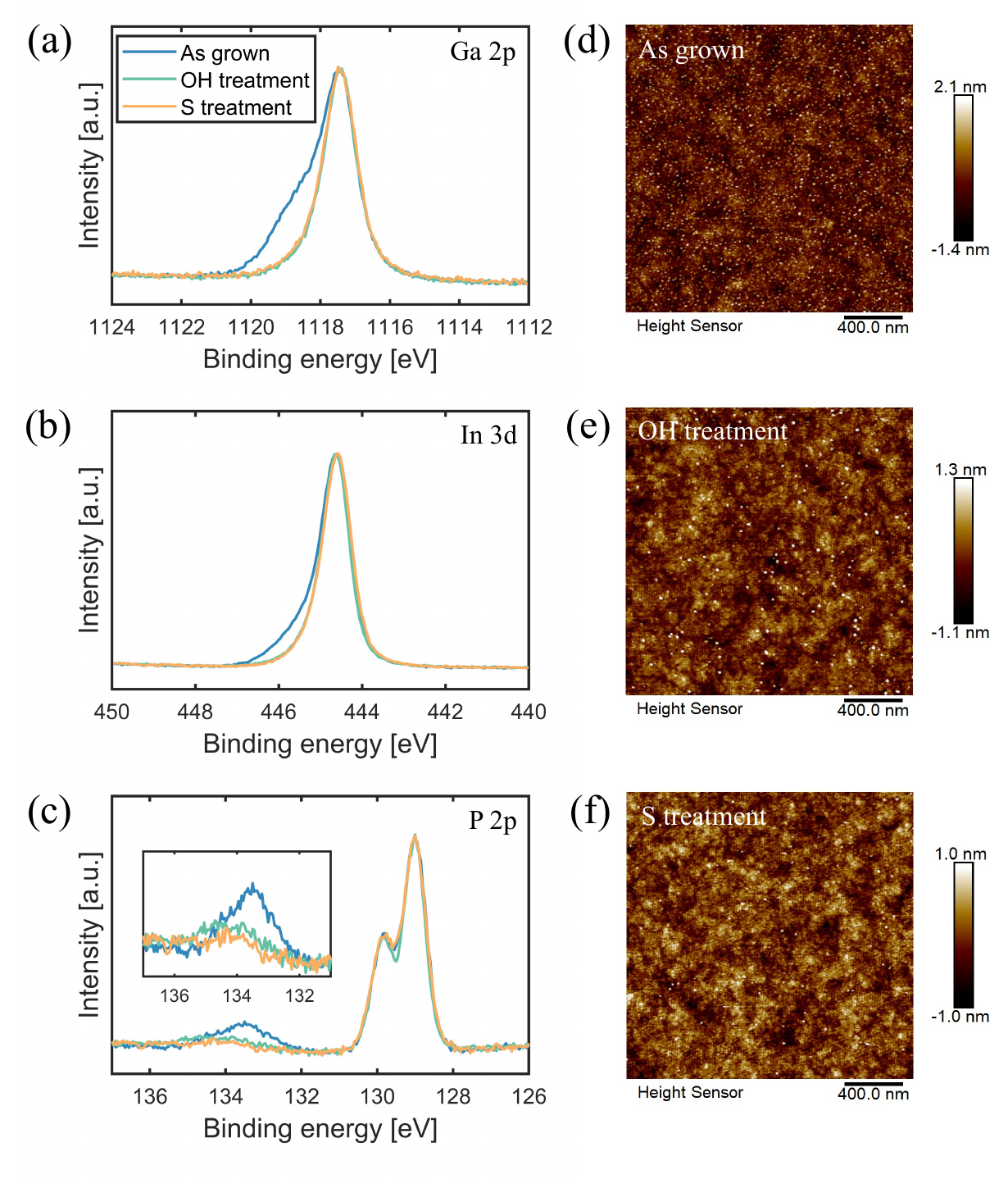}
\caption{
Normalized XPS spectra of (a) Ga $2p$, (b) In $3d$, and (c) P $2p$ under different surface treatments. AFM images of samples (d) as-grown, (e) OH-treated, and (f) S-treated.}
\label{fig:A2}
\end{figure}

In low-loss photonic devices, the effective ``active region'' corresponds to the spatial region with strong modal field overlap, i.e., the top and bottom surfaces for the $780\,\mathrm{nm}$ $\mathrm{TM}_{00}$ and $\mathrm{TM}_{01}$ modes. Consequently, surface states at these interfaces can become a dominant source of optical loss and must be carefully controlled to minimize propagation loss at visible wavelengths where photon energies approach defect levels~\cite{guha2017surface}. The oxidation of InGaP surfaces in air occurs almost instantaneously, and effective surface passivation requires not only dielectric deposition but also prevention of native oxide formation.

\section{Low-Loss InGaP-OI}
Here, we introduce a surface passivation approach based on $(\mathrm{NH}_4)_2\mathrm{S}$ treatment, where sulfur atoms bond to the III--V surface to suppress oxidation and terminate dangling bonds~\cite{wong2023recovering,hasegawa2010surface,jeng2007effective,brennan2011optimisation,laukkanen2021passivation}. We investigate its impact on surface states, surface morphology, propagation loss, and resonator quality factor. The top surfaces of three samples were studied: untreated material (``as-grown''), samples subjected to a $30\,\mathrm{s}$ dip in $10\%$ $\mathrm{NH}_4\mathrm{OH}$ (``OH treatment''), and samples treated with a $10\,\mathrm{min}$ rapid thermal annealing (RTA) in $\mathrm{N}_2$ forming gas followed by a $10\,\mathrm{min}$ dip in $20\%$ $(\mathrm{NH}_4)_2\mathrm{S}$ at $60\,^{\circ}\mathrm{C}$ (``S treatment'').

Surface chemical states were characterized using X-ray photoelectron spectroscopy (XPS). To minimize re-oxidation after treatment, six cycles of atomic layer deposition (ALD) $\mathrm{Al}_2\mathrm{O}_3$ (TMA + $\mathrm{H}_2\mathrm{O}$, $200\,^{\circ}\mathrm{C}$) were deposited immediately following the treatments. The XPS spectra for Ga $2p$, In $3d$, and P $2p$ are shown in Fig.~\ref{fig:A2}(a)--(c), respectively. The binding energy of III--V--oxide bonds is typically higher than that of III--V bonds~\cite{chastain1992handbook}. The OH treatment effectively removes Ga- and In-related oxides, as evidenced by the significant reduction of oxidized components in the Ga $2p$ and In $3d$ spectra, and also reduces P-related oxidation. The P--O bonds are observed at a binding energy of $\sim134\,\mathrm{eV}$, as shown in the zoomed-in inset of Fig.~\ref{fig:A2}(c). In contrast, the S treatment removes Ga- and In-related oxides and provides more effective passivation of P-related surface states. The difference observed in the P $2p$ spectra between the two treatments is subtle, likely because phosphorus oxidation occurs at approximately a monolayer level even after the OH treatment. Meanwhile, a sample subjected only to RTA annealing, without $(\mathrm{NH}_4)_2\mathrm{S}$ treatment, exhibits even more severe surface oxidation. 

\begin{figure}[t]
\centering\includegraphics[width=\textwidth]{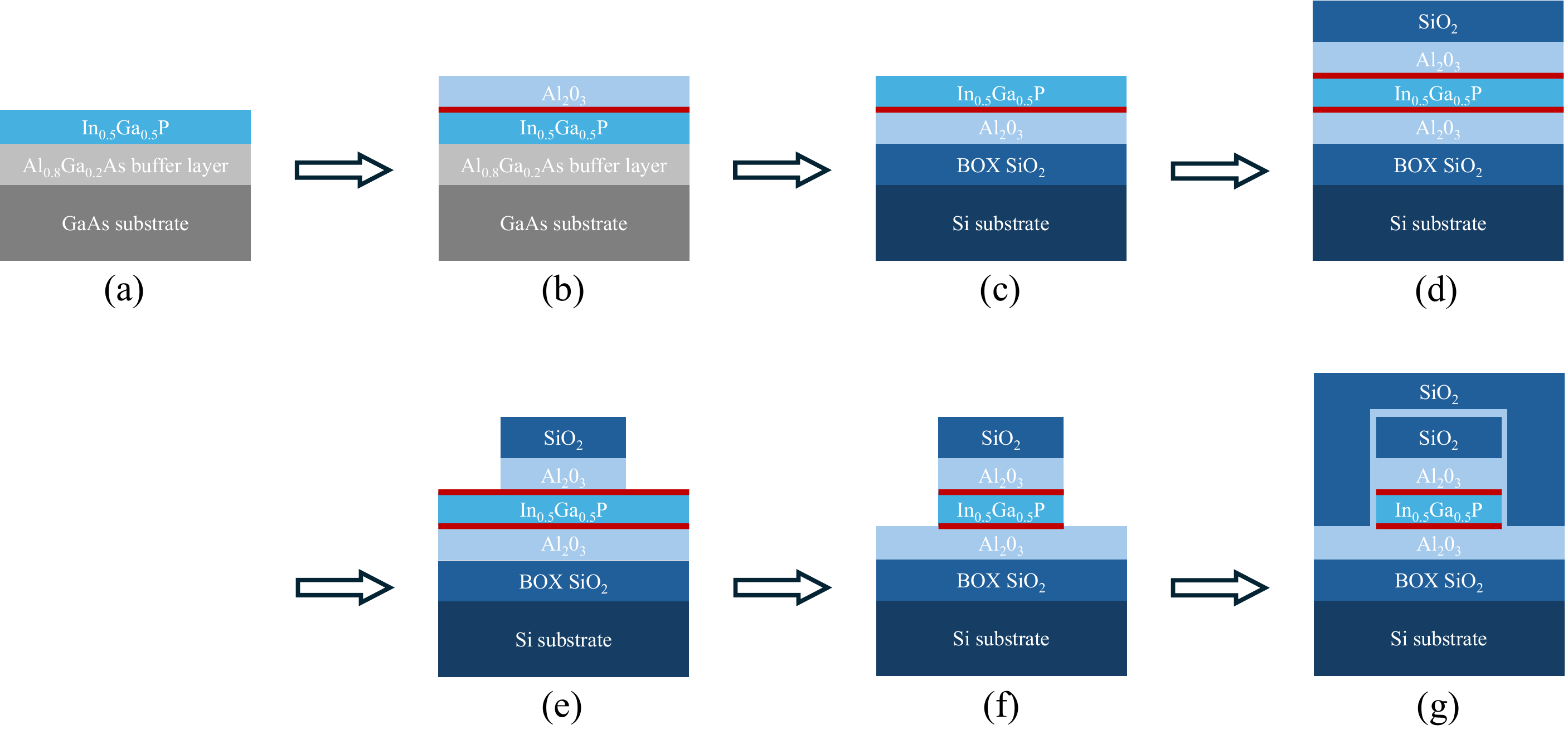}
\caption{
InGaP-OI nanofabrication process flow. 
(a) As-grown InGaP epitaxial. 
(b) S treatment and ALD $\mathrm{Al_2O_3}$ passivation on the InGaP epitaxial. 
(c) Wafer bonding of the InGaP epitaxial to a 4-inch Si wafer with a $3\,\mu$m BOX layer, followed by substrate and buffer layer removal using selective wet etching. 
(d) S treatment of the exposed bottom InGaP surface, followed by ALD deposition of $\mathrm{Al_2O_3}$ passivation and a $\mathrm{SiO_2}$ hard mask. 
(e) Lithography and hard-mask dry etching. 
(f) Waveguide dry etching. 
(g) Sidewall passivation using ALD $\mathrm{Al_2O_3}$ and deposition of a $\mathrm{SiO_2}$ cladding layer.}
\label{fig:A3}
\end{figure}

In addition to chemical passivation, both treatments also improve the surface morphology. As shown in Fig.~\ref{fig:A2}(d)--(f), the root-mean-square (RMS) surface roughness decreases from $0.397\,\mathrm{nm}$ for the as-grown surface to $0.330\,\mathrm{nm}$ after OH treatment and further to $0.295\,\mathrm{nm}$ after S treatment. These reductions in surface oxidation and roughness are expected to suppress both absorption and scattering losses in the microresonators.

The fabrication of microresonators is based on heterogeneous integration of InGaP-OI, as illustrated in Fig.~\ref{fig:A3}. After S treatment and immediate deposition of $5\,\mathrm{nm}$ ALD $\mathrm{Al}_2\mathrm{O}_3$ passivation on the top InGaP surface, the sample is wafer-bonded to a silicon wafer with a $3\,\mu\mathrm{m}$ buried oxide (BOX) layer. The GaAs substrate and $\mathrm{Al}_{0.8}\mathrm{Ga}_{0.2}\mathrm{As}$ buffer layer are subsequently removed by selective wet etching. The same treatment and passivation are then applied to the exposed bottom InGaP surface. An $\mathrm{SiO}_2$ hard mask is deposited prior to waveguide patterning. Waveguides are defined using a DUV stepper with DUV 42P (AR) and UV6 0.8 photoresists. After lithography, the AR layer is removed by $\mathrm{O}_2$ plasma ashing and the hard mask is patterned using a dry etch with $\mathrm{CF}_4/\mathrm{CHF}_3/\mathrm{O}_2$ chemistry. Residual AR is removed using a non-destructive multi-cycle oxidation and downstream plasma ashing process ($\mathrm{H}_2\mathrm{O}_2$ + $\mathrm{O}_2$ plasma), verified by SEM to leave no residue while preserving the hard mask. The InGaP waveguides are then etched using a $\mathrm{Cl}_2/\mathrm{BCl}_3/\mathrm{N}_2$ chemistry, followed by sidewall passivation and deposition of a $2\,\mu\mathrm{m}$ PECVD $\mathrm{SiO}_2$ cladding. 

\section{Linear Characterization}
\begin{figure}[b]
\centering\includegraphics[width=1\columnwidth]{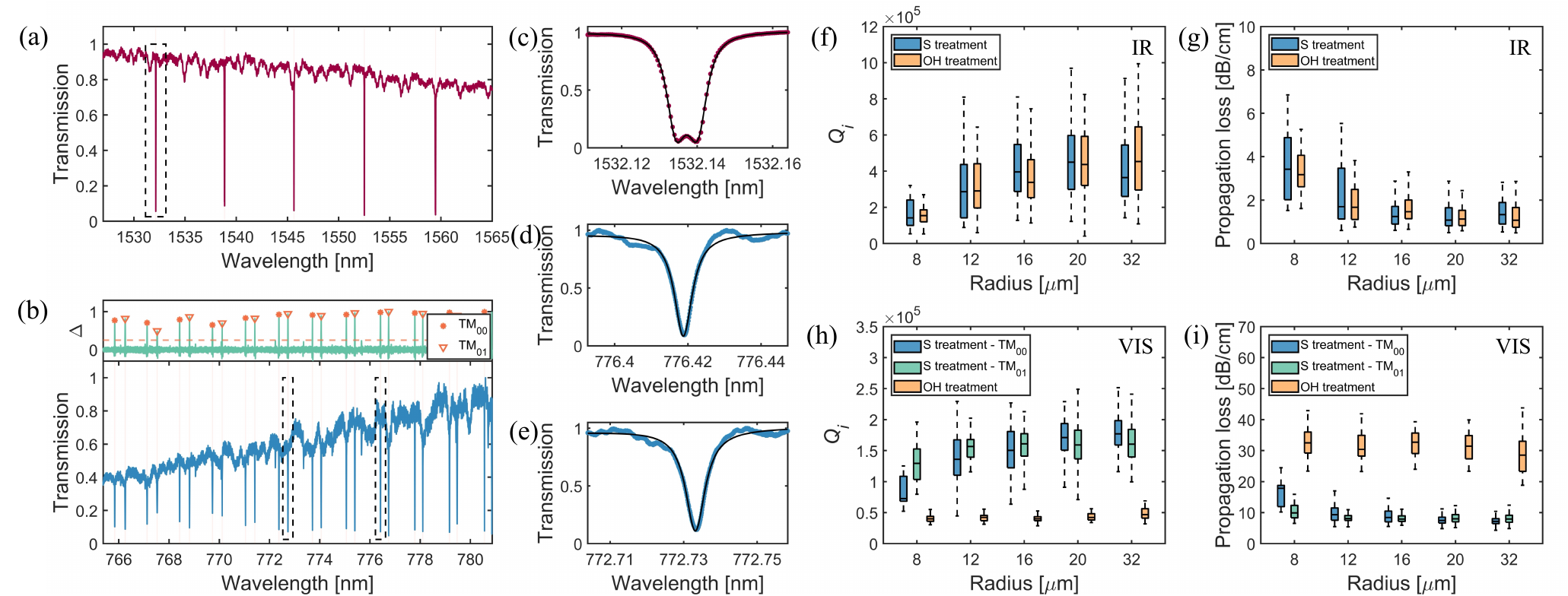}
\caption{
Normalized transmission spectra of a $20\,\mu$m-radius all-pass ring resonators in (a) IR and (b) VIS range. 
Resonance fitting for (c) IR $\mathrm{TE}_{00}$, (d) VIS $\mathrm{TM}_{00}$, and (e) VIS $\mathrm{TM}_{01}$ modes, corresponding to the dashed box in Fig.~4(a)-(b). The IR $\mathrm{TE}_{00}$ resonance is fitted using a split-resonance model~\cite{moresco2013method,de2020mode,pfeiffer2018ultra}, while the VIS $\mathrm{TM}_{00}$ and $\mathrm{TM}_{01}$ resonances are fitted using a standard resonance model~\cite{bogaerts2012silicon}. 
(f-i) Box plots of $Q_i$ and propagation loss for the IR and VIS wavelength ranges for resonators with different radii and surface treatments.}
\label{fig:A4}
\end{figure}

The transmission spectra of $20\,\mu\mathrm{m}$ radius ring resonators with S treatment were measured using lensed fibers coupled to $300\,\mathrm{nm}$ inverse and $2.5\,\mu\mathrm{m}$ flare tapers optimized for the IR and VIS wavelength ranges, respectively, as shown in Fig.~\ref{fig:A4}(a)-(b). The measured facet losses are approximately $3\,\mathrm{dB/facet}$ for the IR and $8\,\mathrm{dB/facet}$ for the VIS in the all-pass configurations.

Multiple transverse magnetic (TM) modes are observed in the VIS wavelength range. To reliably detect all TM resonances, we introduce a mode-detection parameter $\Delta$:

\begin{equation}
\Delta = 1 - \frac{T}{T_{\mathrm{Lowpass}}},
\end{equation}

\noindent where $T$ is the transmission and $T_{\mathrm{Lowpass}}$ is a smoothed transmission spectrum obtained via low-pass filtering. This procedure enhances resonance contrast and enables robust identification of all resonances. The $\mathrm{TM}_{00}$ mode exhibits a slightly larger free spectral range (FSR) than $\mathrm{TM}_{01}$, consistent with simulation, and the mode identification is further confirmed by the corresponding SHG.

Representative resonance fits are shown in Fig.~\ref{fig:A4}(c)--(e) for $\mathrm{TE}_{00}$ ($\sim1532.14\,\mathrm{nm}$, $Q_l = 239.8\mathrm{k}$, $Q_i = 852.0\mathrm{k}$, $\eta_{\mathrm{esc}} = 71.9\%$, loss $= 0.58\,\mathrm{dB/cm}$), $\mathrm{TM}_{00}$ ($\sim776.42\,\mathrm{nm}$, $Q_l = 132.0\mathrm{k}$, $Q_i = 204.6\mathrm{k}$, $\eta_{\mathrm{esc}} = 35.5\%$, loss $= 6.22\,\mathrm{dB/cm}$), and $\mathrm{TM}_{01}$ ($\sim772.73\,\mathrm{nm}$, $Q_l = 120.9\mathrm{k}$, $Q_i = 181.6\mathrm{k}$, $\eta_{\mathrm{esc}} = 33.4\%$, loss $= 7.05\,\mathrm{dB/cm}$). Here, $Q_l$ is loaded quality factor, $Q_i$ is intrinsic quality factor, and $\eta_{\mathrm{esc}}$ is escape efficiency. The corresponding results for OH-treated transmission spectra and resonance fits are provided in the Supplementary Material.

To quantify device performance for OH and S treatments, we statistically analyzed $320$ resonators with varying radii, coupling gaps, and coupling angles for the IR and VIS wavelength ranges, as shown in Fig.~\ref{fig:A4}(f)--(i). In the IR regime, $\sim400\mathrm{k}$ $Q_i$ and $\sim1.25\,\mathrm{dB/cm}$ propagation loss are reliably obtained in microresonators with radii larger than $16\,\mu\mathrm{m}$. Resonators with radii below $16\,\mu\mathrm{m}$ exhibit noticeable degradation of $Q_i$ and increased propagation loss in OH-treated microresonators, and no measurable improvement is observed after S treatment. This behavior is consistent with limited modal overlap with the top and bottom surfaces at IR wavelengths and the relatively low photon energy compared to surface-state bandgaps, indicating that surface absorption is not the dominant loss channel in the IR regime. Instead, bending related loss becomes increasingly significant and dominates as the resonator radius decreases.

In contrast, S treatment produces a significant improvement ($\sim3.5$--$4\times$) in both $Q_i$ and propagation loss in the VIS regime for both $\mathrm{TM}_{00}$ and $\mathrm{TM}_{01}$ modes. Without S treatment, the $Q_i$ and propagation loss remain relatively insensitive to radius ($\sim40\mathrm{k}$ $Q_i$ and $\sim30\,\mathrm{dB/cm}$ propagation loss), suggesting that surface loss dominates over bending related loss. After S passivation, however, a clear radius-dependent loss trend is observed, indicating that surface dangling bonds were the primary loss channel and have been effectively terminated by sulfur atoms. For the $\mathrm{TM}_{00}$ mode, resonators with radii above $20\,\mu\mathrm{m}$ exhibit a median $Q_i$ of $\sim170\,\mathrm{k}$ and a propagation loss of $\sim7\,\mathrm{dB/cm}$, with best values reaching $\sim251\,\mathrm{k}$ and $\sim4.31\,\mathrm{dB/cm}$, respectively. Degraded $Q_i$ and increased loss are observed for resonators with radii below $20\,\mu\mathrm{m}$. We also observe that resonators with larger coupling gaps exhibit slightly lower loss and higher $Q_i$. Because the nonlinear interaction strength in resonators scales inversely with resonator radius, reducing the radius is as important as increasing the quality factor for enhancing nonlinear processes~\cite{boyd2008nonlinear}. Our results indicate that an optimal radius of approximately $20\,\mu\mathrm{m}$ balances strong nonlinear coupling with high $Q_i$ and low propagation loss. Further reduction of the radius provides limited benefit due to the increasing contribution of bending related loss.

For the OH-treated resonators, $\mathrm{TM}_{00}$ and $\mathrm{TM}_{01}$ modes cannot be unambiguously distinguished due to the absence of one mode and FSR fluctuations arising from strong surface-induced loss in most resonators. Nevertheless, the two modes exhibit similar propagation losses because of comparable modal overlap with the waveguide geometry. The corresponding median and best values of $Q_i$ and propagation losses are provided in Table~\ref{tab:T1}. The highest $Q_i$ or lowest propagation loss are indicated in parentheses. In addition, surface-state termination can modify the refractive index, particularly the imaginary component, affecting the FSR, group index, and effective index. A larger FSR ($\sim10\,\mathrm{GHz}$) is observed in resonators after S treatment, as shown in Fig.~\ref{fig:A1}(d). Furthermore, the FSR versus wavelength trend becomes more stable in resonators with S treatment. These results demonstrate that surface passivation not only reduces optical loss and improves the quality factor but also enables a more reliable photonic platform.

Further reduction of the visible wavelength propagation loss could be achieved through several approaches. First, improved surface passivation could further suppress the absorption associated with native oxides, surface states, and defects introduced during wafer bonding and substrate removal. Second, optimization of the waveguide geometry could reduce the modal overlap with the top and bottom InGaP surfaces and may further reduce propagation loss. Finally, optimization of the dry-etching process could reduce plasma-induced damage and sidewall roughness, although the optical mode in the present waveguide geometry exhibits relatively limited overlap with the sidewalls.

\begin{table*}[h]
\centering
\caption{Median and best values of $Q_i$ and propagation loss for different modes with different treatments in the IR and VIS wavelength ranges.}
\renewcommand{\arraystretch}{1.25}
\setlength{\tabcolsep}{5pt}
\resizebox{\textwidth}{!}{%
\begin{tabular}{c|c|c|c|c|c|c|c|c|c|c}
\hline
\multirow{3}{*}{\makecell{Radius\\$[\mu\mathrm{m}]$}}
& \multicolumn{4}{c|}{IR}
& \multicolumn{6}{c}{VIS} \\
\cline{2-11}

& \multicolumn{2}{c|}{\makecell{OH treatment\\TE$_{00}$}}
& \multicolumn{2}{c|}{\makecell{S treatment\\TE$_{00}$}}
& \multicolumn{2}{c|}{\makecell{OH treatment\\TM$_{00\&01}$}}
& \multicolumn{2}{c|}{\makecell{S treatment\\TM$_{00}$}}
& \multicolumn{2}{c}{\makecell{S treatment\\TM$_{01}$}} \\
\cline{2-11}

& $Q_i$ & \makecell{Loss\\$[\mathrm{dB/cm}]$}
& $Q_i$ & \makecell{Loss\\$[\mathrm{dB/cm}]$}
& $Q_i$ & \makecell{Loss\\$[\mathrm{dB/cm}]$}
& $Q_i$ & \makecell{Loss\\$[\mathrm{dB/cm}]$}
& $Q_i$ & \makecell{Loss\\$[\mathrm{dB/cm}]$} \\
\hline

8  & \makecell{1.54$\times10^{5}$ \\(2.70$\times10^{5}$)} & \makecell{3.17\\(1.61)}
   & \makecell{1.42$\times10^{5}$ \\(3.21$\times10^{5}$)} & \makecell{3.42\\(1.52)}
   & \makecell{4.00$\times10^{4}$ \\(5.50$\times10^{4}$)} & \makecell{32.50\\(23.39)}
   & \makecell{7.25$\times10^{4}$ \\(1.25$\times10^{5}$)} & \makecell{17.84\\(10.19)}
   & \makecell{1.29$\times10^{5}$ \\(1.96$\times10^{5}$)} & \makecell{9.89\\(6.52)} \\
\hline

12 & \makecell{2.91$\times10^{5}$ \\(6.44$\times10^{5}$)} & \makecell{1.67\\(0.76)}
   & \makecell{2.87$\times10^{5}$ \\(8.10$\times10^{5}$)} & \makecell{1.70\\(0.60)}
   & \makecell{4.23$\times10^{4}$ \\(5.55$\times10^{4}$)} & \makecell{30.39\\(23.25)}
   & \makecell{1.36$\times10^{5}$ \\(2.29$\times10^{5}$)} & \makecell{9.31\\(5.48)}
   & \makecell{1.57$\times10^{5}$ \\(2.02$\times10^{5}$)} & \makecell{8.06\\(5.42)} \\
\hline

16 & \makecell{3.38$\times10^{5}$ \\(7.45$\times10^{5}$)} & \makecell{1.47\\(0.65)}
   & \makecell{3.96$\times10^{5}$ \\(8.10$\times10^{5}$)} & \makecell{1.24\\(0.61)}
   & \makecell{4.00$\times10^{4}$ \\(5.26$\times10^{4}$)} & \makecell{32.72\\(24.00)}
   & \makecell{1.50$\times10^{5}$ \\(2.27$\times10^{5}$)} & \makecell{8.42\\(5.54)}
   & \makecell{1.61$\times10^{5}$ \\(2.13$\times10^{5}$)} & \makecell{7.92\\(5.91)} \\
\hline

20 & \makecell{4.37$\times10^{5}$ \\(8.24$\times10^{5}$)} & \makecell{1.12\\(0.59)}
   & \makecell{4.50$\times10^{5}$ \\(9.69$\times10^{5}$)} & \makecell{1.08\\(0.50)}
   & \makecell{4.21$\times10^{4}$ \\(5.63$\times10^{4}$)} & \makecell{31.45\\(23.35)}
   & \makecell{1.71$\times10^{5}$ \\(2.29$\times10^{5}$)} & \makecell{7.47\\(4.83)}
   & \makecell{1.59$\times10^{5}$ \\(2.49$\times10^{5}$)} & \makecell{7.99\\(5.14)} \\
\hline

32 & \makecell{4.53$\times10^{5}$ \\(9.94$\times10^{5}$)} & \makecell{1.07\\(0.49)}
   & \makecell{3.64$\times10^{5}$ \\(9.13$\times10^{5}$)} & \makecell{1.33\\(0.54)}
   & \makecell{4.67$\times10^{4}$ \\(6.89$\times10^{4}$)} & \makecell{28.54\\(18.84)}
   & \makecell{1.77$\times10^{5}$ \\(2.51$\times10^{5}$)} & \makecell{7.16\\(4.31)}
   & \makecell{1.60$\times10^{5}$ \\(2.41$\times10^{5}$)} & \makecell{7.94\\(4.86)} \\
\hline

\end{tabular}%
}
\label{tab:T1}
\end{table*}

\section{Entangled-Photon Pair Generation}
With the linear optical performance of the $20\,\mu\mathrm{m}$ ring resonators demonstrated, we next examine the nonlinear optical response, beginning with entangled photon-pair generation via SPDC. The corresponding wavelengths for the two strong nonlinear coupling conditions, $\Delta m=\pm2$, with ring widths ranging from $1.025$ to $1.175\,\mu\mathrm{m}$ are experimentally determined from the SHG wavelengths, as shown in Fig.~\ref{fig:A1}(e). A detailed analysis of quasi-phase matching, energy conservation, and SHG intensity is provided in the Supplementary Material. The large azimuthal mode numbers indicate that quasi-phase matching occurs near the modal phase-matching wavelength, i.e., around $1100\,\mathrm{nm}$, rather than the $1300\,\mathrm{nm}$ predicted by FDE simulations. This discrepancy is likely caused by the weak confinement of the VIS mode and the inaccurate refractive index of InGaP at the top and bottom surfaces of the resonators. 

\begin{figure}[t]
\centering\includegraphics[width=\textwidth]{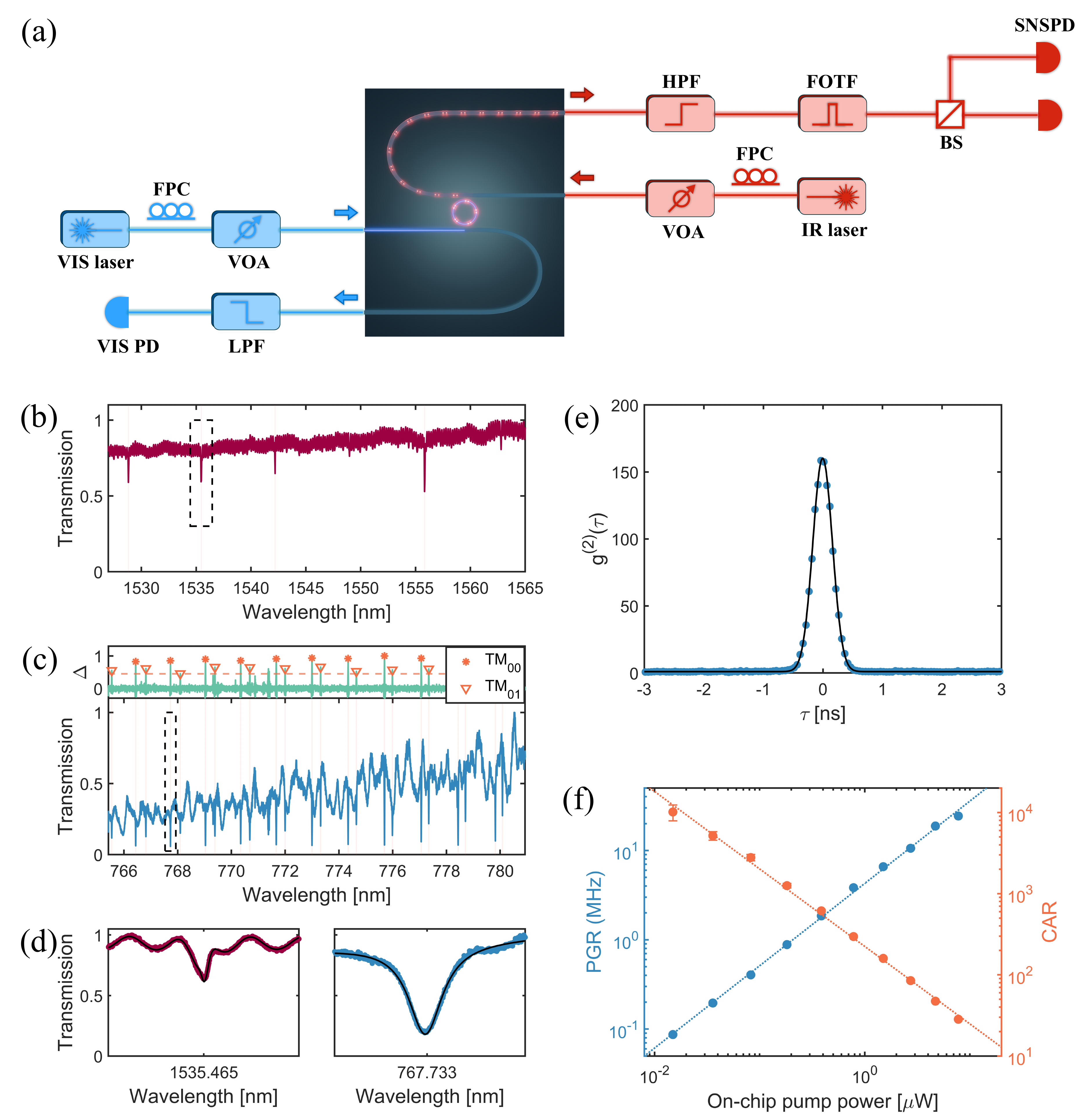}
\caption{
(a) Schematic of the degenerate entangled-photon pair generation measurement setup. FPC, fiber polarization controller; VOA, variable optical attenuator; LPF: low-pass filter; HPF: high-pass filter; BS: beam splitter; SNSPD: superconducting nanowire single-photon detector, PD: photodetector.
Transmission spectrum of a $20\,\mu$m-radius add-drop ring resonator designed for phase matching between the (b) IR $\mathrm{TE}_{00}$ mode and (c) the VIS $\mathrm{TM}_{00}$ mode. 
(d) Normalized transmission spectrum and resonance fitting for the IR $\mathrm{TE}_{00}$ and VIS $\mathrm{TM}_{00}$ modes, corresponding to the dashed box in Fig.~5(b)-(c). The IR $\mathrm{TE}_{00}$ resonance is fitted using a split-resonance model~\cite{moresco2013method,de2020mode,pfeiffer2018ultra}.
(e) Second-order correlation function of the degenerate SPDC photon pairs measured at an on-chip pump power of $1.50\,\mu$W. The binwidth was set to 50 ps. (f) Degenerate PGR and CAR as a function of pump power. The PGR in bus waveguide is calculated by integrating the coincidence counts within the $\pm 3\sigma$ integration window of the Gaussian fit to the coincidence histogram. CAR is calculated by $g^{(2)}(0)-1$~\cite{signorini2020chip}. Error bars represent the uncertainty estimated from shot noise.}
\label{fig:A5}
\end{figure}

The schematic of SPDC measurement setup is shown in Fig.~\ref{fig:A5}(a). Here, we present an add-drop structure for strong photon-pair generation. The ring width is $1050\,\mathrm{nm}$, with a $200\,\mathrm{nm}$ coupling gap to an $810\,\mathrm{nm}$-wide IR bus waveguide and a $250\,\mathrm{nm}$ gap to a $450\,\mathrm{nm}$-wide VIS bus waveguide. To accurately achieve doubly resonant conditions, microheaters are fabricated above the resonator to fine-tune the refractive index and align the IR and VIS resonances. The microheaters are easy to fabricate, highly compact, and stable during testing with no measurable drift in the resonance wavelengths. The transmission spectra and resonance fits after fine thermal tuning are shown in Fig.~\ref{fig:A5}(b)--(d). For the IR resonance at $\sim1535.47\,\mathrm{nm}$, $Q_l = 33.6\mathrm{k}$, $Q_i = 367.9\mathrm{k}$, $\eta_{\mathrm{esc}} = 90.9\%$, and propagation loss $= 1.35\,\mathrm{dB/cm}$. For the VIS resonance at $\sim767.74\,\mathrm{nm}$, $Q_l = 73.7\mathrm{k}$, $Q_i = 102.7\mathrm{k}$, $\eta_{\mathrm{esc}} = 28.2\%$, and propagation loss $= 12.6\,\mathrm{dB/cm}$. The $Q_i$ at both resonances is slightly reduced compared with the all-pass resonators, primarily due to the interaction of both bus waveguides with the ring. Photon-pair generation is performed by pumping the device with a hopping-free continuous-wave (CW) VIS laser (New Focus TLB-6700). Light is coupled onto the chip using fiber arrays at both wavelengths. The generated photon pairs are detected by superconducting nanowire single-photon detectors (SNSPDs) after applying the optimal thermal current to achieve resonance alignment. Two long-pass filters with a total extinction ratio of $100\,\mathrm{dB}$ are used to suppress leaked VIS pump photons. 

Both degenerate and nondegenerate photon pairs are generated in the microresonator. The pump wavelength and thermal tuning current are  optimized by maximizing the degenerate photon-pair count, measured using two etalon-based fiber-optic tunable filters (FOTFs, Agiltron) with a total extinction ratio of 50 dB and a bandwidth of 1 nm. A WaveShaper (Finisar 4000A) was first used to filter the degenerate SPDC resonance during the pump wavelength and thermal current tuning. The FOTF was then added and fine-tuned to maximize the transmission at the degenerate wavelength. After the optimization, the WaveShaper was removed before the photon-pair measurements.
Figure~\ref{fig:A5}(e) shows the measured second-order correlation of the generated degenerate SPDC photon pairs, exhibiting a pronounced bunching peak. The photon-pair generation rate (PGR) and coincidence-to-accidental ratio (CAR) under different on-chip pump powers are presented in Fig.~\ref{fig:A5}(f). A degenerate PGR slope of $4.27\,\mathrm{MHz}/\mu\mathrm{W}$ is obtained in the bus waveguide, reaching $6.59\,\mathrm{MHz}$ PGR at a CAR of $159$. Fitting the second-order correlation histogram can independently extract an intracavity degenerate PGR slope of $4.85\,\mathrm{MHz}/\mu\mathrm{W}$~\cite{guo2017parametric}, as described in the Supplementary Material. The selected nondegenerate signal-idler pair at 1528.83 nm and 1542.20 nm exhibits a PGR slope of $3.06\,\mathrm{MHz}/\mu\mathrm{W}$ in the bus waveguide. Details of the measurements are provided in the Supplementary Materials. These results demonstrate highly efficient quantum light generation in compact, low-loss InGaP-OI integrated waveguides enabled by cavity-enhanced nonlinear interactions.

\section{Second Harmonic Generation}
\begin{figure}[h]
\centering\includegraphics[width=\textwidth]{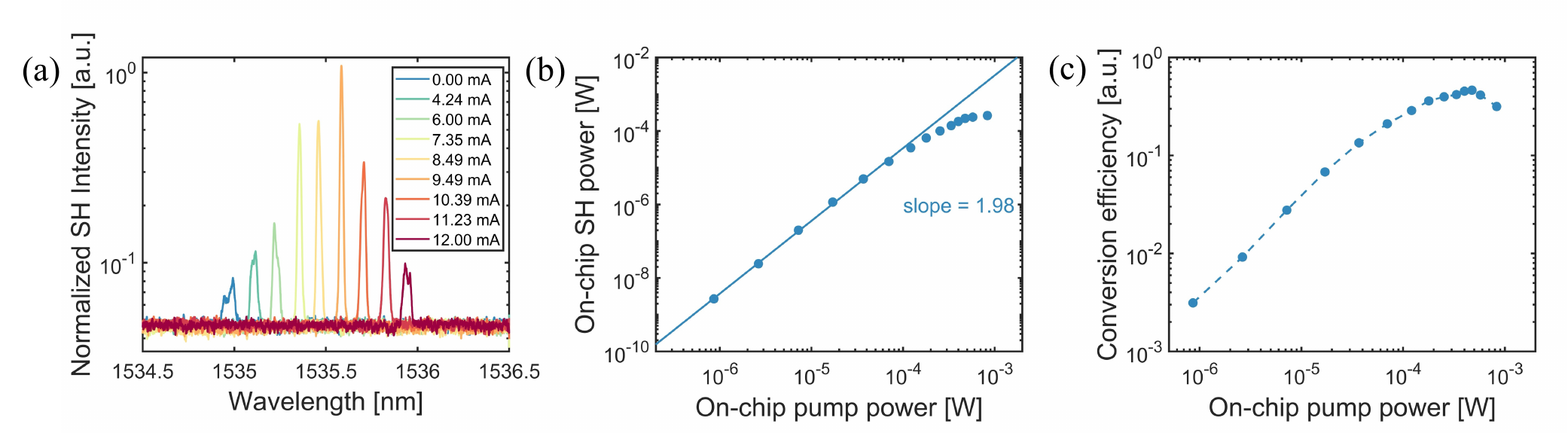}
\caption{
(a) Normalized SHG intensity under different thermal current biases. The SHG spectrum redshifts with increasing thermal current. 
(b) On-chip SHG power as a function of pump power, confirming a quadratic dependence. 
(c) Conversion efficiency as a function of on-chip pump power.}
\label{fig:A6}
\end{figure}

Although the add-drop structure is not optimized for SHG due to the strong escape efficiency in the IR range, significant SHG is still observed in the $20\,\mu\mathrm{m}$ resonator configuration. The same resonator is pumped using a mode-hopping-free CW IR laser (Toptica DLC Pro), and the current applied to the microheaters is tuned until the maximum SHG intensity is achieved. The shorter-wavelength peak corresponds to the VIS resonance, while the longer-wavelength peak corresponds to the IR resonance when no thermal current is applied. As the thermal current increases, a clear trend is observed where improved resonance alignment is achieved and the VIS resonance redshifts faster than the IR resonance, as shown in Fig.~\ref{fig:A6}(a). When the thermal current is further increased, the resonance alignment and SHG intensity decrease. With the optimal thermal current applied, the SHG signal is collected and measured using a power meter (Newport 2936-R). The relationship between SHG intensity and IR pump power is shown in Fig.~\ref{fig:A6}(b). A fitted slope of $1.98$ is obtained, confirming the quadratic dependence of SHG on pump power, and an on-chip SHG efficiency of $3.01\times10^{5}\,\%/\mathrm{W}$ is achieved, which is comparable to suspended InGaP rings with smaller radii~\cite{zhao2022ingap,akin2024ingap}. Furthermore, a peak conversion efficiency (CE) of $46.3\%$ is observed at $0.48\,\mathrm{mW}$ on-chip pump power, as shown in Fig.~\ref{fig:A6}(c). 

A comparison of the state-of-the-art SHG performance of $\chi^{(2)}$ microring resonators is presented in Table~2. The  high-performance InGaP-OI platform presented in this work simultaneously offers high SHG efficiency, high conversion efficiency, and compatibility with heterogeneous integration .  

An attractive future direction is the realization of cascaded $\chi^{(2)}$ nonlinear processes on the InGaP-OI platform. By combining efficient SHG with SPDC in appropriately engineered microresonators, the generated second-harmonic field could directly pump an integrated SPDC stage, enabling compact sources of entangled photon pairs and squeezed light ~\cite{zhang2021high,shi2024efficient}. The low propagation loss, high-quality-factor resonators, and experimentally verified quasi-phase-matching conditions demonstrated in this work provide a promising foundation for realizing such fully integrated cascaded nonlinear photonic circuits.

\begin{table*}[h]
\centering
\caption{Comparison of state-of-the-art SHG performance in integrated $\chi^{(2)}$ microring resonators.}
\renewcommand{\arraystretch}{1.25}
\setlength{\tabcolsep}{5pt}
\resizebox{\textwidth}{!}{%
\begin{tabular}{c|c|c|c|c|c}
\hline
\makecell{Platform}
& \makecell{SHG efficiency\\$[\%/\mathrm{W}]$}
& \makecell{Maximum \\SH power $[\mathrm{mW}]$}
& \makecell{Maximum \\CE $[\%]$}
& \makecell{Pump power at\\maximum CE $[\mathrm{mW}]$}
& \makecell{SH wavelength\\$[\mathrm{nm}]$} \\
\hline

PP-LNOI~\cite{lu2020toward}
& $5.0\times10^{6}$
& 0.02
& --
& --
& 752 \\
\hline

Suspended InGaP membrane~\cite{akin2024ingap}
& $4.4\times10^{5}$
& --
& --
& --
& 767 \\
\hline

\textbf{InGaP-OI (this work)}
& \textbf{\boldmath $3.0\times10^{5}$}
& \textbf{\boldmath 0.26}
& \textbf{\boldmath 46.3}
& \textbf{\boldmath 0.48}
& 768 \\
\hline

Epitaxial AlN-on-sapphire~\cite{bruch201817}
& $1.7\times10^{4}$
& 11
& 11
& 35
& 775 \\
\hline

GaP-OI~\cite{logan2018400}
& $4.0\times10^{2}$
& 0.025
& 0.84
& --
& 775 \\
\hline

4H-SiCOI~\cite{lukin20204h}
& $3.6\times10^{2}$
& 0.001
& --
& --
& 778 \\
\hline

\end{tabular}%
}
\label{tab:SHG_comparison}
\end{table*}

\section{Outlook}
In summary, we demonstrate a heterogeneously integrated InGaP-OI nonlinear photonics platform with low propagation loss for efficient frequency conversion and entanglement generation. Through modal analysis and sulfur-based passivation of InGaP surface dangling bonds, we significantly reduce surface roughness and suppress surface absorption losses, achieving a $\sim3.5$--$4\times$ reduction in propagation loss and intrinsic quality factors approaching $\sim250\mathrm{k}$ in the visible wavelength range. Statistical analysis of resonators with varying radii further identifies the optimal radius for maximizing nonlinear interaction while avoiding significant bending related loss. Using a $20\,\mu\mathrm{m}$ radius doubly resonant microresonator, we demonstrate a degenerate photon-pair generation rate of $4.27\,\mathrm{MHz}/\mu\mathrm{W}$ with a high coincidence-to-accidental ratio, as well as a second-harmonic generation efficiency of $3.01\times10^{5}\,\%/\mathrm{W}$. With optimized coupling conditions (critical coupling for both resonances) and achievable quality factors of $\sim969\,\mathrm{k}$ and $\sim229\,\mathrm{k}$ for the IR and VIS modes, respectively, an on-chip SHG efficiency of $2.20\times10^{7}\,\%/\mathrm{W}$ and a photon-pair generation rate (PGR) of $93.58\,\mathrm{MHz}/\mu\mathrm{W}$ can be achieved. These results establish InGaP-OI as a promising platform for integrated nonlinear and quantum photonics, enabling scalable on-chip frequency conversion and entangled photon sources.

\begin{backmatter}

\bmsection{Funding} 
This work was supported by the Defense Advanced Research Projects Agency (Award No. D24AC00166-00) and the NSF (Quantum Foundry Grant No. DMR-1906325, CAREER Grant No. 2045246, and NRT Grant No. 2152201). L.T. and L.W. acknowledge support from the NSF Graduate Research Fellowship Program.

\bmsection{Disclosures}
X.L. and G.M. are inventors on a provisional patent application related to the (NH$_4$)$_2$S surface treatment for low-loss InGaP microresonators reported in this work.

\bmsection{Data Availability Statement}
The data supporting the findings of this work are available from the corresponding author upon reasonable request.

% \bmsection{Supplemental document}
% \label{sec:Supplement 1}
% See Supplement 1 for supporting content.

\end{backmatter}

%%%%%%%%%%%%%%%%%%%%%%% References %%%%%%%%%%%%%%%%%%%%%%%%%

\bibliography{references}

\end{document}